# New test for the Hubble law

Jacek Chołoniewski ★
*Astronomical Observatory of the Warsaw University, Aleje Ujazdowskie 4, 00-478 Warsaw, Poland*

**ABSTRACT**
The new, simple method is proposed for testing the Hubble law using redshifts and magnitude taken from magnitude–limited sample of galaxies. The power law relation between redshift and distance have been assumed ($cz = r^p$). The obtained results: $p = 1.21 \pm 0.04$ (for CfA) and $p = 0.95 \pm 0.04$ (for ESO/LV) confirm linear, Hubble expansion. The proposed test does not take into account inhomogeneites in space distriubution of galaxies.

## 1 INTRODUCTION

The problem of testing validity of the Hubble law using as observational database redshifts and magnitude of galaxies taken from magnitude–limited sample have been addressed in several papers of Nicoll, Segal and colaborators (see Nicoll & Segal 1980, Nicoll & Segal 1983, Segal 1989, Segal, Nicoll & Wu 1994 and references therein) and recently by Strauss & Willick (1995) and Koranyi & Strauss (1995).

We raise again this problem by proposing a new, simple test.

## 2 THE TEST

Let us consider an observed joint distribution function of luminosity ($L$) and distance ($r$): $N(L, r)$ for a flux–limited sample of galaxies. According to Chołoniewski (1995) this distribution can be expressed as:

$$N(L, r) = \Phi(L)\, r^2\, I\!\left(\frac{L}{r^2} \geq f_{lim}\right) \tag{1}$$

where the indicator function $I(.)$, which describes here observational flux cut–off, is defined as:

$$I(statement) = \begin{cases} 1 & \text{if the } statement \text{ is true} \\ 0 & \text{if the } statement \text{ is false} \end{cases} \tag{2}$$

and where $\Phi(L)$ denotes luminosity function. Equation (1) is valid under following assumptions:

(i) the number density of galaxies is contant,
(ii) luminosity ($L$) and space coordinates (**r**) of galaxies are statistically independent,
(iii) the space has Euclidean properties.

★ E-mail: jch@sirius.astrouw.edu.pl

$$f = \frac{L}{r^2}. \tag{3}$$

Let us compute now the joint distribution function of luminostiy ($L$) and flux ($f$): $N(L, f)$. It can be done by the transformation of variables in equation (1) from ($L, r$) to ($L, f$), what gives:

$$N(L, f) = \Phi(L)\, L^{3/2}\, f^{-5/2}\, I(f \geq f_{lim}) \tag{4}$$

Equation (4) is in agreement with a "fundamental theorem" introduced by Neyman & Scott (1974). Equation (4) shows that luminosity and flux of galaxies in a flux–limited sample are statistically independent. If so, the conditional distribution function is equivalent to the marginal distribution function:

$$N(L \mid f) = \Phi(L)\, L^{3/2}. \tag{5}$$

Let us assume now (in agreement with all authors quoted in Section 1) that redshifts and distances are connected by the relation:

$$cz = r^p. \tag{6}$$

According to equation (3) the redshift $cz$ can be expressed as a function of $L$ and $f$:

$$cz = L^{p/2}\, f^{-p/2} \tag{7}$$

what will be used in computing the expectation value of redshift ($cz$) for a given flux ($f$):

$$E(cz \mid f) = \int_0^\infty L^{p/2}\, f^{-p/2}\, N(L \mid f)\, dL \tag{8}$$

and what finally gives:

$$E(cz \mid f) = const\, f^{-p/2}. \tag{9}$$

The magnitude version of equation (9) is:

$$E(\log cz \mid m) = 0.2\, p\, m + const, \tag{10}$$

where:

$$m = -2.5 \log(f) + const. \tag{11}$$

The left hand side of equation (10) defines a so-called regression curve of the first type (see Fisz 1963, Section 3.7). Since, as we have shown, the shape of this curve should be linear we can compute (according to Fisz 1963, Section 3.8) the slope coeficient in equation (10), which is equal to $0.2p$, using the least square linear regression (regression of the second type).

Now, we can formulate our test for the Hubble law: if the Hubble law holds the relation between the average logarithm of redshift and magnitude is linear with a slope equal to 0.2.

The important and somewhat surprising feature of the result expressed in equation (10) is that it is completely insensitive to the shape of the luminosity function of galaxies, so, the slope of the relation expressed in equation (10) is strictly equal to the expected slope for the ideal case when galaxies had the same luminosities (luminosity function equal to the delta function). It should be emphasized however, that this feature exists *only* for the power law relation between redshift and distance – see equations (6), (7) and (8).

## 3  APPLICATION TO THE DATA

As an observational database we used two magnitude–limited samples of galaxies: CfA (Davies & Huchra 1982) and ESO/LV (Lauberts & Valentijn 1989) – see Chołoniewski (1995) for details. Raw data for both samples are presented in Figs 1 and 2.

Let us compare now theoretical predictions expressed in equation (10) with observational data. In order to do it we have computed the average logarithm of redshift for a given magnitude: $< \log cz \mid m >$. The resultant averages are presented in Figs 3 and 4 as one standard deviation error bars connected by broken line. They lie approximately along a straight line what confirms the validity of equation (10). This agreement allow us to use simple linear regression in order to compute the parameter $p$ as described in Section 2. The resultant straight lines are presented in Figs 3 and 4. According to equation (10) their slopes are equal to $0.2p$ what corresponds to: $p = 1.21 \pm 0.04$ (for CfA) and $p = 0.95 \pm 0.04$ (for ESO/LV). No wonder that the agreement with Hubble law ($p = 1$) is better for ESO/LV than for CfA. The CfA sample is dominated by one neighbouring supercluster (Virgo) which distorts significantly two of our assumptions: homogeneity of the galaxy space distribution and velocity field described by a power law (equation 6).

## 4  DISCUSSION

Our result is in agreement with Strauss & Willick (1995) who obtained that $p \approx 1$ (what corresponds to the Hubble law) and with disagreement with Nicoll & Segal (see Section 1 for references) who advocated $p \approx 2$ (what corresponds to the Lundmark law). Certainly, our result that $p \approx 1$ is in agreement with many other publications which are based on various distance indicators for individual galaxies.

One possible explanation of Nicoll & Segal's result that $p \approx 2$ is given by Koranyi & Strauss (1995). They reported that the test used by Nicoll & Segal is "almost independent of $p$". This fact, however, can not be taken as the ultimate explanation since Nicoll & Segal applied their method to *many* different samples of extragalactic objects *always* arriving to $p \approx 2$. We will propose another explanation in a future work.

We show in this paper that the assumption that the number density of galaxies is constant implies, when confronted with observational data, that the Hubble law holds ($p \approx 1$). Strauss & Willick (1995) presented the argumentation which goes in the opposite direction (but leads to the same conclusion): they assume different values of $p$ and found that just for $p \approx 1$ the observed (in terms of IRAS 1.2 Jy sample) Universe is approximate homogeneous.

The assumption about homogeneity of galaxy space distribiution which has been applied throughout this paper, although acceptable as a first approximation, is not true – galaxies are clustered. The challenging task for future work in this subject arises here: to develop a test which would not need homogeneity assumption.

## 5  CONCLUSION

We have shown that if the Hubble law holds the relation between the average logarithm of redshift and magnitude should be linear with a slope equal to 0.2. This simple fact can be applied for testing the validity of the Hubble law using any magnitude–limited sample of galaxies with known redshifts. We have applied the proposed method to CfA and ESO/LV samples of galaxies. Our results confirms the validity of the Hubble law.

## ACKNOWLEDGMENTS

This work was partly supported by KBN grant to the Warsaw University Observatory.

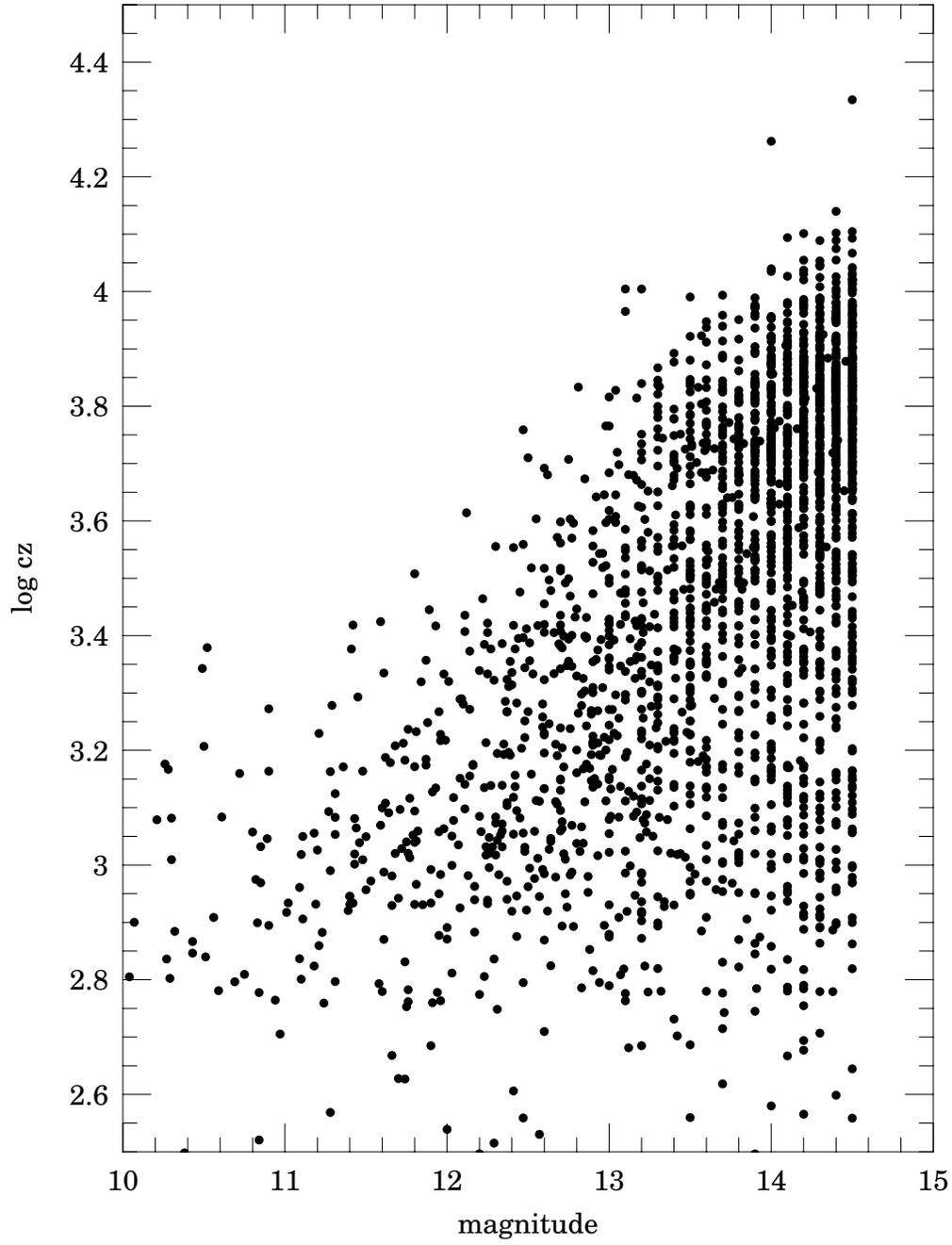

**Figure 1.** The magnitude – logarithm of redshift diagram for 2373 galaxies taken from CfA sample.

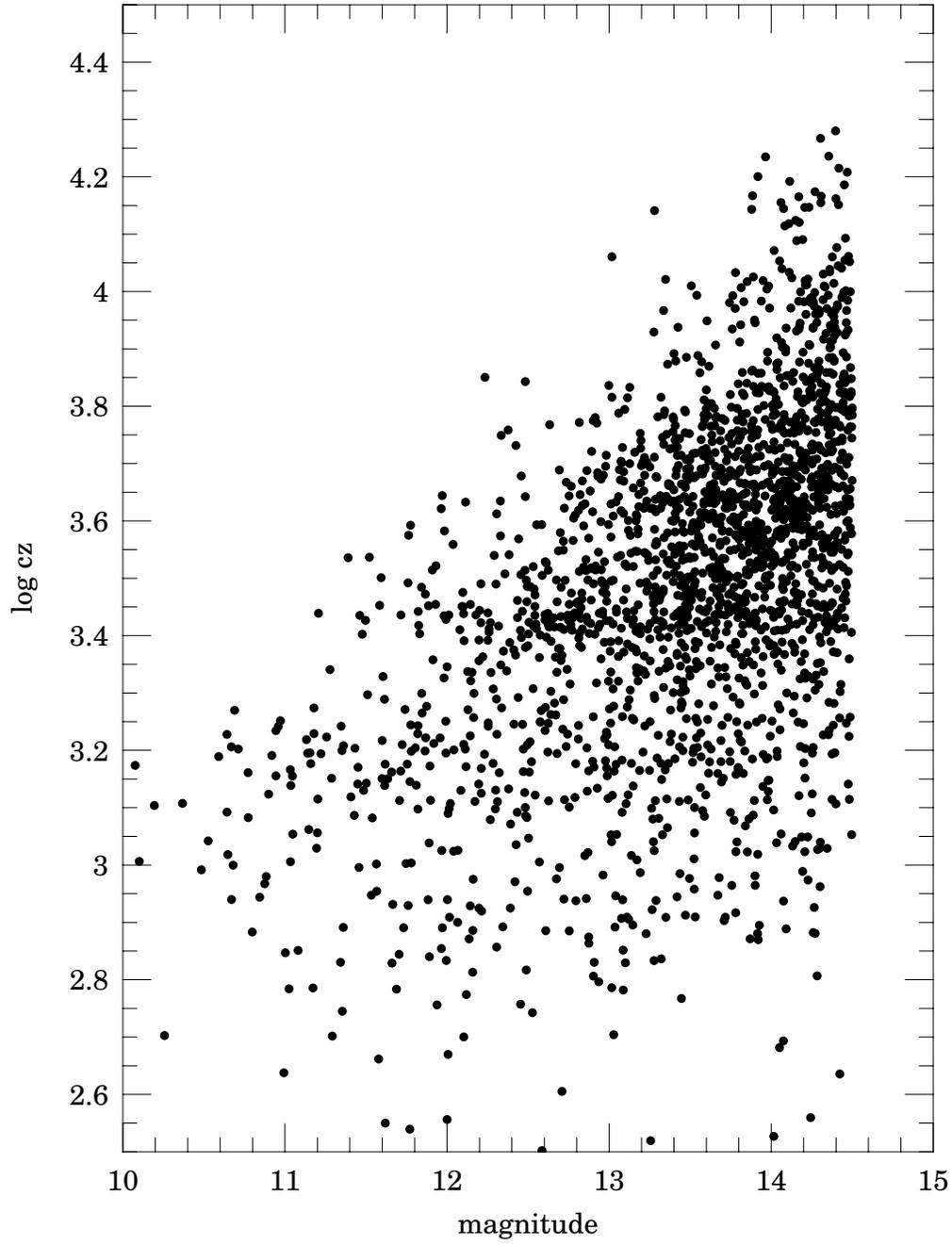

**Figure 2.** The same as Fig. 1 but for for 2062 galaxies taken from ESO/LV sample.

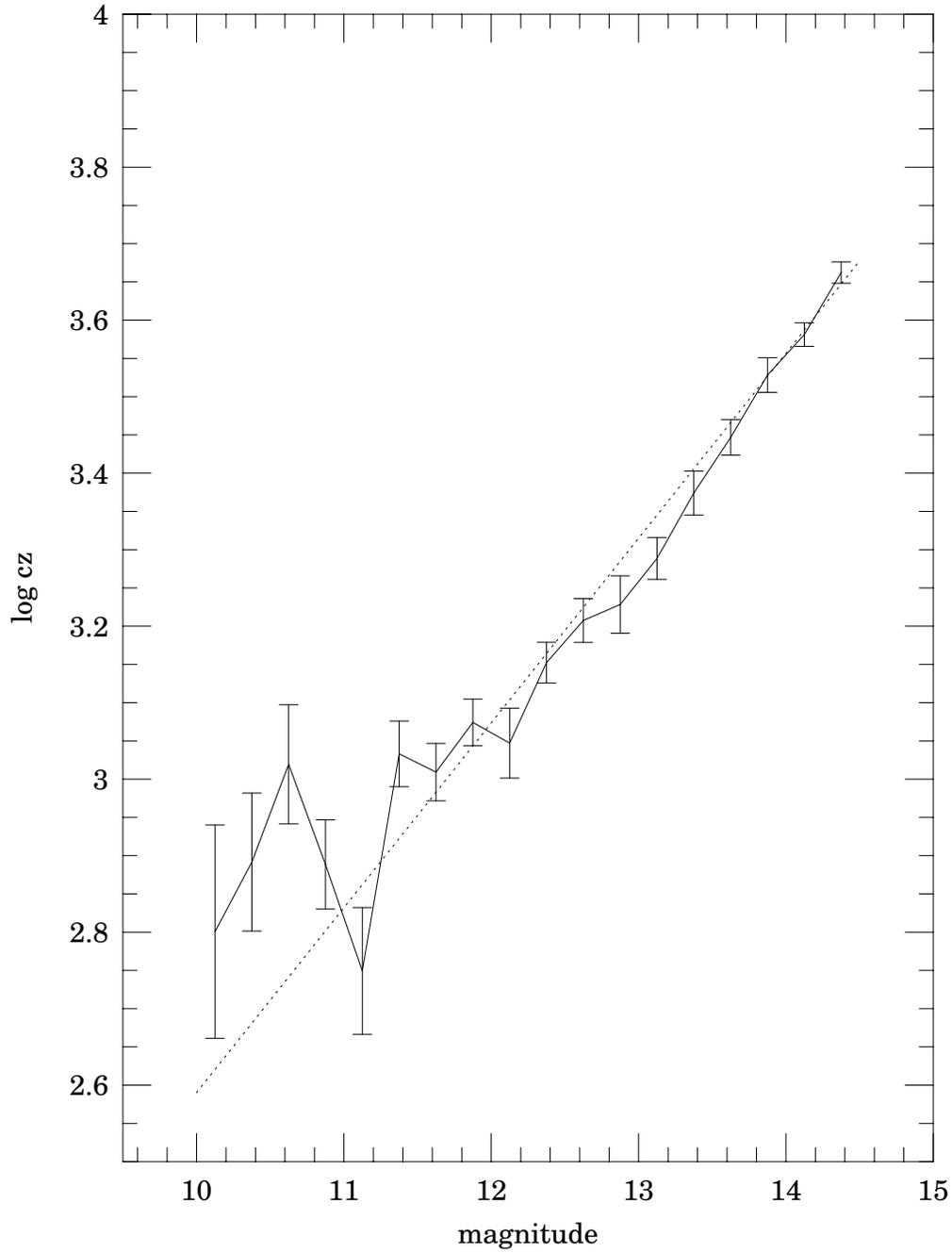

**Figure 3.** The average logarithm of redshift as a function of magnitude for CfA sample of galaxies. The averages are represented by one standard deviation error bars connected by a broken line. The dotted line represents linear least square regression. Its slope is close to 0.2 – the value expected for the Hubble law.

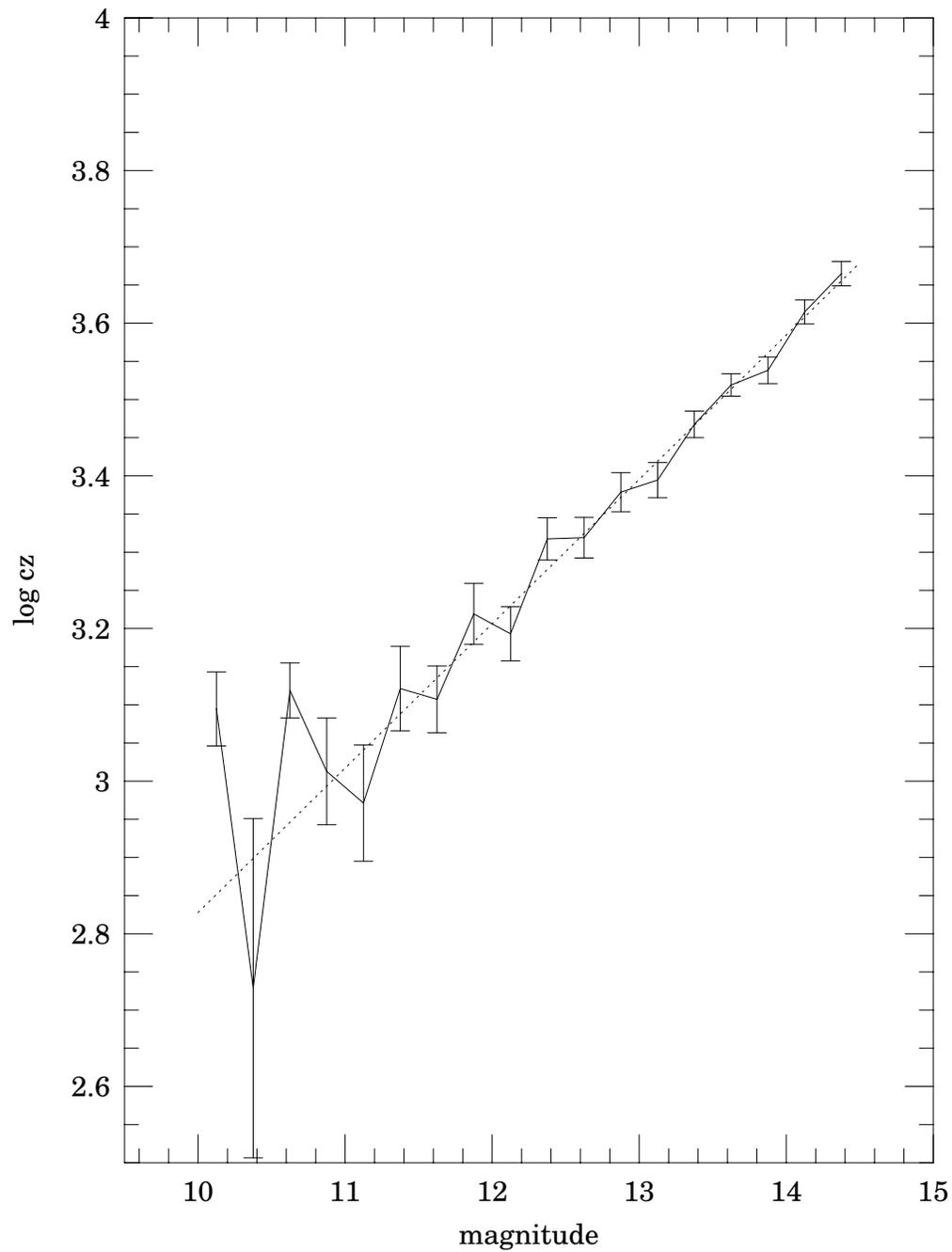

**Figure 4.** The same as Fig. 3 but for ESO/LV sample of galaxies.